\documentclass[11pt]{article}
\pdfoutput=1
\usepackage[centertags]{amsmath}
\usepackage[square, comma, sort&compress,numbers]{natbib}
\usepackage{array,multirow}
\numberwithin{equation}{section}
\usepackage{amssymb,amsfonts}
\usepackage{graphicx}
\usepackage{color}
\usepackage{mathtools}

\usepackage{epsfig}
\usepackage{bbold}

\usepackage{wrapfig}
\usepackage{float}
\usepackage{soul}
\usepackage{tkz-euclide}

\usepackage{tikz,pgf}
\usetikzlibrary{shapes}
\usetikzlibrary{calc}
\usetikzlibrary{decorations.pathmorphing}
\usetikzlibrary{decorations.pathreplacing,shapes.misc}
\usetikzlibrary{positioning}
\usetikzlibrary{arrows}
\usetikzlibrary{decorations.markings}
\usetikzlibrary{shadings}

\usetikzlibrary{intersections}

\def\be{\begin{equation}}
\def\ee{\end{equation}}
\def\ba{\begin{array}}
\def\ea{\end{array}}

\def\dps{\displaystyle}

\def\1{\tilde{1}}
\def\2{\tilde{2}}
\def\3{\tilde{3}}

\newdimen\tableauside\tableauside=1.0ex
\newdimen\tableaurule\tableaurule=0.4pt
\newdimen\tableaustep
\def\phantomhrule#1{\hbox{\vbox to0pt{\hrule height\tableaurule
width#1\vss}}}
\def\phantomvrule#1{\vbox{\hbox to0pt{\vrule width\tableaurule
height#1\hss}}}
\def\sqr{\vbox{%
  \phantomhrule\tableaustep

\hbox{\phantomvrule\tableaustep\kern\tableaustep\phantomvrule\tableaustep}%
  \hbox{\vbox{\phantomhrule\tableauside}\kern-\tableaurule}}}
\def\squares#1{\hbox{\count0=#1\noindent\loop\sqr
  \advance\count0 by-1 \ifnum\count0>0\repeat}}
\def\tableau#1{\vcenter{\offinterlineskip
  \tableaustep=\tableauside\advance\tableaustep by-\tableaurule
  \kern\normallineskip\hbox
    {\kern\normallineskip\vbox
      {\gettableau#1 0 }%
     \kern\normallineskip\kern\tableaurule}%
  \kern\normallineskip\kern\tableaurule}}
\def\gettableau#1 {\ifnum#1=0\let\next=\null\else
  \squares{#1}\let\next=\gettableau\fi\next}

\newcommand{\bref}[1]{\textbf{\ref{#1}}}

\def\cF{\mathcal{F}}

\def\cH{\mathcal{H}}

\def\cL{\mathcal{L}}

\def\cO{\mathcal{O}}

\def\cV{\mathcal{V}}

\numberwithin{equation}{section} \makeatletter
\@addtoreset{equation}{section}

\def\be{\begin{equation}}
\def\ee{\end{equation}}
\def\ba{\begin{array}}
\def\ea{\end{array}}

\def\dps{\displaystyle}

\def\ba{\begin{array}}
\def\ea{\end{array}}

\def\dps{\displaystyle}

\def\kk{\text{,}}

\newmuskip\pFqmuskip
\newcommand*\pFq[6][8]{%
  \begingroup 
  \pFqmuskip=#1mu\relax
  \mathcode`\,=\string"8000
  \begingroup\lccode`\~=`\,
  \lowercase{\endgroup\let~}\pFqcomma
  {}_{#2}F_{#3}{\left[\genfrac..{0pt}{}{#4}{#5};#6\right]}%
  \endgroup
}
\newcommand{\pFqcomma}{\mskip\pFqmuskip}

\newmuskip\LpFqmuskip
\newcommand*\LpFq[6][8]{%
  \begingroup 
  \pFqmuskip=#1mu\relax
  \mathcode`\,=\string"8000
  \begingroup\lccode`\~=`\,
  \lowercase{\endgroup\let~}\pFqcomma
  {}_{}F_{}{\left[\genfrac..{0pt}{}{#4}{#5};#6\right]}%
  \endgroup
}

\newmuskip\Ftmuskip
\newcommand*\Ft[6][8]{%
  \begingroup 
  \pFqmuskip=#1mu\relax
  \mathcode`\,=\string"8000
  \begingroup\lccode`\~=`\,
  \lowercase{\endgroup\let~}\pFqcomma
  F_{2}{\left[\genfrac..{0pt}{}{#4}{#5};#6\right]}%
  \endgroup
}
\newmuskip\Ftmuskip
\newcommand*\FK[6][8]{%
  \begingroup 
  \pFqmuskip=#1mu\relax
  \mathcode`\,=\string"8000
  \begingroup\lccode`\~=`\,
  \lowercase{\endgroup\let~}\pFqcomma
  F_{K}{\left[\genfrac..{0pt}{}{#4}{#5};#6\right]}%
  \endgroup
}

\usepackage{jheppub}
\makeatletter
\def\@fpheader{\vspace{-.1cm}}
\makeatother

\title{Global torus blocks in the necklace channel}

\author[a]{Mikhail\ Pavlov}

\affiliation[a]{I.E. Tamm Department of Theoretical Physics, \\P.N. Lebedev Physical
Institute,\\ Leninsky ave. 53, 119991 Moscow, Russia}

\emailAdd{pavlov@lpi.ru}

\abstract{ We continue studying of global conformal blocks on the torus in a special (necklace) channel. Functions of such multi-point blocks are explicitly found under special conditions on the blocks' conformal dimensions. We have verified that these blocks satisfy the Casimir equations, which were derived in previous studies.}

\makeatletter
\def\@fpheader{\vspace{-.1cm}}
\makeatother

\begin{document}

\maketitle
\flushbottom

\flushbottom

\section{Introduction}

In CFT$_2$, main observables are correlation functions of local operators. The global and local symmetries of the theory impose restrictions on these correlators. In particular, multi-point correlators of primary operators can be decomposed into conformal blocks \cite{Belavin:1984vu}, which are determined only by the symmetries of the theory. Conformal blocks associated with local symmetry algebras such as Virasoro (or extended $W_N$ algebras \cite{Fateev:1987vh, Fateev:2011qa}) have been studied on the sphere \cite{Belavin:1984vu, Fitzpatrick:2014vua, Hijano:2015rla, Perlmutter:2015iya} and on the torus \cite{Hadasz:2009db, Kraus:2017ezw, Gobeil:2018fzy}.

The \textit{global} blocks (associated with $sl(2, \mathbb{R})$ algebra) on the sphere were considered in \cite{Ferrara:1974ny, Dolan:2000ut, Dolan:2003hv, Osborn:2012vt}. Furthermore, these blocks arise in  the large-$c$ limit\footnote{This regime drastically simplifies the Virasoro block functions; moreover, large-$c$ conformal blocks are dual to the Witten diagrams in AdS \cite{Hijano:2015qja, Hijano:2015zsa}.} from the Virasoro blocks in different regimes \cite{Harlow:2011ny, Fitzpatrick:2014vua, Fitzpatrick:2015zha} . On the torus, the study was initiated in \cite{Hadasz:2009db}, where the $1$-point block was computed. For $N\geq 2$ correlation functions, there are multiple conformal blocks. For instance, $N$-pt global torus blocks in the OPE channel \cite{Kraus:2017ezw} were explicitly found to be a product of the $(N+2)$-point global block on the sphere and the $1$-point torus block.

In this work, we focus on another type of global torus blocks - so-called \textit{necklace} blocks \cite{Alkalaev:2017bzx, Alkalaev:2022kal}. The crucial point is that various mixed channel blocks (where inserting projectors and taking OPE are combined) can be computed using the necklace channel block (see section 2 in \cite{Alkalaev:2022kal} for details). In this sense, the necklace block corresponding only to the insertion of projectors and the OPE block which is responsible for pure OPE decompositions generate blocks with an arbitrary channel topology. Nevertheless, functions of necklace blocks are not known in the closed form but it was shown that they are subject of Casimir equations \cite{Alkalaev:2022kal}.  Here, we directly construct the $N$-point necklace block in terms of a particular $(N+2)$-point comb channel block \cite{Rosenhaus:2018zqn} and find the functions of the necklace blocks after imposing special conditions on the conformal dimensions to be relatively simple polynomial functions, which allows to reveal their properties and singularities. We also show that these blocks satisfy the Casimir equations mentioned above.

 This paper is organized as follows. In Section \bref{sec:tc} we introduce the torus blocks in the necklace channel and present a relation between the $N$-pt necklace channel block and the $(N+2)$-pt comb channel block. Section \bref{sec:ebf} contains examples of necklace block functions. By imposing certain conditions on the comb channel block, we calculate the $N$-pt blocks in the necklace channel. Section \bref{sec:CE} is devoted to checking that the found blocks satisfy the Casimir equations. Section \bref{sec:cn} summarizes our results and introduces future directions. Appendix \bref{sec:nvo} describes details about necklace and OPE global conformal blocks. In Appendix \bref{mex} we list necklace block functions that supplement the narration in Section \bref{sec:ebf}.

\section{Torus global conformal blocks}
\label{sec:tc}
\paragraph{$sl(2)$ representation theory and correlation functions on the torus.} We consider CFT$_2$ with the local $sl(2, \mathbb{C}) \simeq sl(2, \mathbb{R}) \oplus \overline{sl(2, \mathbb{R})}$ symmetry on the torus which is characterized by the modular parameter $q$.
The generators of the holomorphic part are denoted by $L_{m}$ and they satisfy the commutation relations
\be
\label{sl}
[L_{m}, L_{n}] = (m-n) L_{m+n}, \qquad m, n = 0, \pm 1.
\ee
For the algebra  $\overline{sl(2, \mathbb{R})}$ we have same relations in terms of generators $\bar L_m$ and $[L_{m}, \bar L_n] = 0$. A highest weight state is defined as follows
\be
L_0 | h \rangle = h | h \rangle, \qquad L_1 | h \rangle = 0,
\ee
and the corresponding Verma module $\cV_{h}$ is spanned by the descendants states
\be
|m, h\rangle = L^m_{-1} | h \rangle.
\ee
We will often use $sl(2, \mathbb{C})$ states $| m, m', h, \tilde h \rangle  =| m, h \rangle \otimes | m', \bar h \rangle $ and associated Verma modules $\cV_{h, \bar h}$. The projector on a Verma module  $\cV_{h_i}$ takes the form
\be
\label{pro}
\mathbb{P}_{i} = \sum^{\infty}_{m=0}\frac{|m, h_i\rangle \langle m, h_i|}{m!(2h_i)_m},
\ee
where $(2h_i)_m = \Gamma(2h_i +m)/\Gamma(2h_i)$ is the (rising) Pochhammer symbol. Here the standard conjugation $(L_{-1})^{\dagger} = L_1$ is assumed and for \eqref{pro} we have
\be
\label{propro}
\sum_{h_i \in D}  \mathbb{P}_{i} =  1, \qquad \mathbb{P}^2_{i} = \mathbb{P}_{i},
\ee
where in the first formula the sum is carried over the chiral part of the spectrum $D$. Primary operators $\cO_{h, \bar h} (z, \bar z)$ can be introduced via the operator-state correspondence
\be
\label{OSC}
| h, \tilde h \rangle = \lim_{z, \bar z\rightarrow 0} \cO_{h, \tilde h} (z, \bar z)|0\rangle.
\ee
In addition, the algebra $sl(2, \mathbb{R})$  acting on primaries can be realized in terms of differential operators with respect to the variable $z$
\be
\label{DO}
\mathcal{L}_P = \left (z^{P+1} \partial + z^{P} h (P+1) \right), \qquad P=0, \pm 1.
\ee
In what follows, we will also use $\mathcal{L}^{(k)}_P$ to denote the operators \eqref{DO} with respect to the variable $z_k$.  Note that the commutation relations for such operators differ from \eqref{sl} by the sign on the right side.
\vspace{-2mm}
\paragraph{Torus matrix elements.} The $N$-point correlation function of primary operators $\cO_{i}(z_i, \bar z_i)$  with conformal dimensions $(h_i, \bar h_i)$ on the torus is given by
\be
\label{cf}
\langle \cO_{1} (z_1, \bar z_1)...\cO_{N}(z_N, \bar z_N) \rangle_{_\mathbb{T}} = \operatorname{Tr}_{\cH} \left ( q^{L_0} \bar q^{\bar L_0} \cO_{1} (z_1, \bar z_1)...\cO_{N}(z_N, \bar z_N)\right),
\ee
where $\operatorname{Tr}_{\cH}$ stands for the sum over all Verma modules with weights $(h_{\alpha}, \bar h_{\alpha})$, i.e.
\be
\label{correlator}
\ba{c}
\vspace{2mm}
\langle \cO_{1} (z_1, \bar z_1)...\cO_{N}(z_N, \bar z_N) \rangle_{_\mathbb{T}} = \dps \sum_{h_{\alpha}, \bar h_{\alpha}\in D} \sum^{\infty}_{m, m'=0} \frac{q^{h_{\alpha}+m}  \bar q^{\bar h_{\alpha} +  m'}}{m! m'! (2h_{\alpha})_m (2\bar h_{\alpha})_{m'}}
\\
\times  \langle m, m', h_{\alpha}, \bar h_{\alpha}| \cO_{1} (z_1, \bar z_1)...\cO_{N}(z_N, \bar z_N) |m,  m', h_{\alpha}, \bar h_{\alpha} \rangle. ~~~~~~
\ea
\ee
 Here we apply $L_0 | m, h_a \rangle = (h_{\alpha} +m) | m, h_a \rangle$ to isolate a $q$-dependence. The $N$-point function satisfies Ward's identities associated with the $u(1)\oplus u(1)$ symmetry on the torus
\be
\sum^N_{i=1} \mathcal{L}^{(i)}_{0} \langle \cO_{1} (z_1, \bar z_1)...\cO_{N}(z_N, \bar z_N) \rangle_{_\mathbb{T}} = 0, \qquad \sum^N_{i=1} \mathcal{\bar L}^{(i)}_{0} \langle\cO_{1} (z_1, \bar z_1)...\cO_{N}(z_N, \bar z_N) \rangle_{_\mathbb{T}} = 0.
\ee

The matrix element in the second line of \eqref{correlator} can be expressed in terms of the $(N+2)$-point correlation function on the sphere with two additional operators with dimensions $(h_{\alpha}, \bar h_{\alpha})$. Indeed, using \eqref{OSC} the differential realization of the descendant states reads
\be
\ba{c}
\label{in-out}
|m,  m', h_{\alpha}, \bar h_{\alpha} \rangle = \dps \lim_{z, \bar z\rightarrow 0} \partial^m \bar \partial^m \cO_{\alpha} (z, \bar z)|0\rangle,
\\
\\
\langle m, m', h_{\alpha}, \bar h_{\alpha}| = \langle 0| \dps \lim_{z, \bar z\rightarrow 0}  \partial^m \bar \partial^m \left( \bar z^{-2\bar h_{\alpha}}  z^{-2 h_{\alpha}} \dps  \cO_{\alpha} ( z^{-1}, \bar z^{-1}) \right).
\ea
\ee
where we use the differential realization of the operator $L_{-1}$ as $\partial$. Thus, the expression for the torus matrix element appearing in
\eqref{correlator} can be cast into the following form
\be
\label{mf}
\ba{c}
 \langle m, m', h_{\alpha}, \bar h_{\alpha}| \cO_{1} (z_1, \bar z_1)...\cO_{N}(z_N, \bar z_N) |m,  m', h_{\alpha}, \bar h_{\alpha} \rangle =  \\
 \\
 \lim \limits_{\substack{%
\\
z_0 \to 0\\
z_{N+1} \to 0}} \partial^m_{0} \partial^m_{N+1} \bar \partial^{m'}_{0} \bar \partial^{m'}_{N+1} \left ( z_0^{-2h_{\alpha}}  \bar{z}_0^{-2\bar h_{\alpha}} \langle \cO_{\alpha} ( z_0^{-1}, \bar z_0^{-1}) \cO_{1} (z_1, \bar z_1)...\cO_{N}(z_N, \bar z_N) \cO_{\alpha}(z_{N+1}, \bar z_{N+1}) \rangle \right).
\ea
\ee
The illustrative example of the relation above is a computation of the norm $\big| |m,  m', h_{\alpha}, \bar h_{\alpha} \rangle\big|^2$. Here, we have a $2$-pt function $\langle \cO_{\alpha} \cO_{\alpha}\rangle$ of primary operators with dimensions $(h_{\alpha}, \tilde h_{\alpha})$ in the second line of \eqref{mf}, so
\be
\ba{c}

\vspace{3mm}

\big| |m,  m', h_{\alpha}, \bar h_{\alpha} \rangle\big|^2 = \lim \limits_{\substack{%
\\
z_0, \bar z_0 \to 0\\
z_1, \bar z_1 \to 0}} \partial^m_{0} \partial^m_{1} \bar \partial^{m'}_{0} \bar \partial^{m'}_{1} \left( (1- z_1 z_0)^{-2 h_{\alpha}} (1- \bar z_1 \bar z_0)^{-2 \bar h_{\alpha}} \right)  \\

= m! m'! (2h_{\alpha})_m  (2 \bar{h}_{\alpha})_{m'}\;,
\ea
\ee
where the Pochhammer symbols and factorials come from taking derivatives with respect to $z_1$ and $z_0$, respectively.
\paragraph{The torus block in the necklace channel.}

The correlation functions \eqref{cf} can be decomposed into conformal blocks in two significantly different ways, which we discuss in Appendix \bref{sec:nvo} in details. The first one corresponds to taking the OPE between operators in \eqref{cf} and the second one to inserting projectors \eqref{pro} on the Verma modules with the set of intermediate dimensions. The former was elaborated in \cite{Kraus:2017ezw} where $N$-pt OPE blocks were found in a closed form. Here we focus on the  latter (necklace) channel for which the correlation function \eqref{cf} can be cast into the following form
\be
\label{decomposition}
\langle \cO_{1} (z_1, \bar z_1)...\cO_{N}(z_N, \bar z_N) \rangle_{_\mathbb{T}} = \sum_{\{ \tilde h_i, \bar{\tilde h}_i\} \in D} \mathfrak{Ch}_{h,\tilde{h}} \cF_{N} (q, \mathbf{z}|h, \tilde h, h_{\alpha}) \bar \cF_{N} (\bar q, \bar{\mathbf{z}}|\bar h, \bar{\tilde h}, \bar{h}_{\alpha}),
\ee
where $\cF_{N} (q, \mathbf{z}|h, \tilde h, h_{\alpha})$ is a $N$-point \textit{necklace} channel block on the torus which is parameterized by external dimensions of primary operators $h_j, ~ j= 1, ..., N$, intermediate dimensions $\tilde h_i, ~ i = 0, ..., (N-2)$ and $h_{\alpha}$ appearing in \eqref{cf} so altogether they are succinctly denoted by a set $\{h, \tilde h, h_{\alpha}\}$. The notation $\mathbf{z}$ (or $\bar{\mathbf{z}}$) stands for insertion points of the primary operators $\mathbf{z} = \{z_1, ..., z_N\}$. $\mathfrak{Ch}_{h,\tilde{h}}$ represents as the product of structure constants
\be
\label{tc}
\mathfrak{Ch}_{h,\tilde{h}} = C_{ h_{\alpha}, h_1, \tilde h_{0}} \prod^{N-3}_{i=0} C_{\tilde h_{i}, h_{i+2}, \tilde h_{i+1}} C_{ \tilde h_{N-2}, h_N, h_{\alpha}},  \qquad C_{h_k, h_l, h_m} = \langle h_{k}, \bar{h}_{k}  | \cO_{l} (1, 1) | h_{m}, \bar{h}_{m} \rangle.
\ee
Since the torus matrix element \eqref{mf} can be written through the $(N+2)$-pt correlation function on the sphere one can insert $(N-1)$ projectors which results in expansion into conformal blocks  for this correlation function. The corresponding blocks (denoted by $G_{N+2}$) are comb channel blocks \cite{Rosenhaus:2018zqn}, so from \eqref{mf} we find
\be
\label{maintool}
\ba{c}
\cF_{N} (q, \mathbf{z}|h, \tilde h, h_{\alpha}) = \dps \sum^{\infty}_{m=0} \frac{q^{m+h_{\alpha}} A^{(N+2)}_m ( \mathbf{z}|h, \tilde h, h_{\alpha})}{m! (2h_{\alpha})_m}, \\
\\
A^{(N+2)}_m ( \mathbf{z}|h, \tilde h, h_{\alpha}) = \lim \limits_{\substack{%
\\
z_0 \to 0\\
z_{N+1} \to 0}} \partial^m_{_0} \partial^m_{_{N+1}} \left( z^{-2 h_{\alpha}}_0 G_{N+2} (1/z_0,\mathbf{z}, z_{N+1}|h, \tilde h, h_{\alpha})\right),
\ea
\ee
where  $G_{N+2} (z_0,\mathbf{z}, z_{N+1}|h, \tilde h, h_{\alpha})$ is the $(N+2)$-point comb channel block with external dimensions $(h_{\alpha}, h_1,..., h_{N}, h_{\alpha})$ and intermediate dimensions $( \tilde h_0,..., \tilde h_{N-2})$.

\begin{figure}[H]
\centering
\begin{tikzpicture}[scale=0.8]
 \tkzDefPoint(0,0){OO}
  \tkzDefPoint(1.0,0){B}
    \tkzDefPoint(2,0){A}
      \tkzDefPoint(2.25,0){C}


 \draw[color = black, line width  = 0.9] (0, -1) -- (0, -2.0);
 \draw[color = black, line width  = 0.9] (0, 1) -- (0, 2.0);

  \draw[color = black, line width  = 0.9] (-0.8515, -0.5) -- (-1.73, -1.0);
  \draw[color = black, line width  = 0.9] (-0.8515, 0.5) -- (-1.73, 1.0);

  \tkzDrawCircle[line width  = 1.5, color = red](OO,B)


\fill[black](-1.73, -1.0) circle (0.3mm);
\fill[black](-1.73, 1.0)circle (0.3mm);
\fill[black](0, -2) circle (0.3mm);
\fill[black](0, 2)  circle (0.3mm);

  \draw  (2, 0.0) node {$=$};


\fill[black](-1.73, -1.0) circle (0.3mm);
\fill[black](-1.73, 1.0)circle (0.3mm);
\fill[black](0, -2) circle (0.3mm);
\fill[black](0, 2)  circle (0.3mm);

  \tkzDefPoint(4.5,0){O2}
  \tkzDefPoint(5.5,0){B2}
  \tkzDefPoint(4.5,1){OA1}
  \tkzDefPoint(4.5,-1){OA2}

  \draw[color = black, line width  = 0.9] (3.6585, -0.5) -- (2.77, -1.0);
  \draw[color = black, line width  = 0.9] (3.6585, 0.5) -- (2.77, 1.0);
 \draw[color = black, line width  = 0.9] (4.5, -1) -- (4.5, -2.0);
 \draw[color = black, line width  = 0.9] (4.5, 1) -- (4.5, 2.0);

 \tkzDrawArc[rotate,color=red,line width  = 1.5](O2,OA1)(180)
 \tkzDrawArc[rotate,color=black,line width  = 0.9](O2,OA1)(-70)
 \tkzDrawArc[rotate,color=black,line width  = 0.9](O2,OA2)(70)

 \fill[black](2.77, -1.0) circle (0.3mm);
\fill[black](2.77, 1.0)circle (0.3mm);
\fill[black](4.5, -2) circle (0.3mm);
\fill[black](4.5, 2)  circle (0.3mm);

   \draw  (6.5, 0.0) node {$+$};

\tkzDefPoint(9,0){O3}
  \tkzDefPoint(10.0,0){B3}
  \tkzDefPoint(9,1){OA3}
  \tkzDefPoint(9,-1){OA4}

  \draw[color = black, line width  = 0.9] (8.1585, -0.5) -- (7.27, -1.0);
  \draw[color = black, line width  = 0.9] (8.1585, 0.5) -- (7.27, 1.0);
 \draw[color = black, line width  = 0.9] (9, -1) -- (9, -2.0);
 \draw[color = black, line width  = 0.9] (9, 1) -- (9, 2.0);

  \tkzDrawArc[rotate,color=red,line width  = 1.5](O3,OA3)(180)
 \tkzDrawArc[rotate,color=black,line width  = 0.9](O3,OA3)(-70)
 \tkzDrawArc[rotate,color=black,line width  = 0.9](O3,OA4)(70)

  \fill[black](7.27, -1.0) circle (0.3mm);
\fill[black](7.27, 1.0)circle (0.3mm);
\fill[black](9, -2) circle (0.3mm);
\fill[black](9, 2)  circle (0.3mm);

 \fill[red](9.95, -0.3) circle (0.65mm);
 \fill[red](9.95, 0.3) circle (0.65mm);

   \draw  (11.0, 0.0) node {$+$};

 \tkzDefPoint(13.5,0){O4}
  \tkzDefPoint(13.5,1){OA5}
  \tkzDefPoint(13.5,-1){OA6}

  \draw[color = black, line width  = 0.9] (12.6585, -0.5) -- (11.77, -1.0);
  \draw[color = black, line width  = 0.9] (12.6585, 0.5) -- (11.77, 1.0);
 \draw[color = black, line width  = 0.9] (13.5, -1) -- (13.5, -2.0);
 \draw[color = black, line width  = 0.9] (13.5, 1) -- (13.5, 2.0);

 \tkzDrawArc[rotate,color=red,line width  = 1.5](O4,OA5)(180)
 \tkzDrawArc[rotate,color=black,line width  = 0.9](O4,OA5)(-70)
 \tkzDrawArc[rotate,color=black,line width  = 0.9](O4,OA6)(70)

  \fill[black](11.77, -1.0) circle (0.3mm);
\fill[black](11.77, 1.0)circle (0.3mm);
\fill[black](13.5, -2) circle (0.3mm);
\fill[black](13.5, 2)  circle (0.3mm);

 \fill[red](14.45, -0.3) circle (0.65mm);
  \fill[red](14.41, -0.41) circle (0.65mm);
 \fill[red](14.45, 0.3) circle (0.65mm);
 \fill[red](14.41, 0.41) circle (0.65mm);

    \draw  (16, 0.0) node {$+ ~~~...$};


\fill[black](-1.5, 0)  circle (0.2mm);
\fill[black](-1.445, 0.4)  circle (0.2mm);
\fill[black](-1.445, -0.4)  circle (0.2mm);

\fill[black](-1.072, 1.0)  circle (0.2mm);
\fill[black](-0.829, 1.25)  circle (0.2mm);
\fill[black](-0.538, 1.4)  circle (0.2mm);

\fill[black](-1.072, -1.0)  circle (0.2mm);
\fill[black](-0.829, -1.25)  circle (0.2mm);
\fill[black](-0.538, -1.4)  circle (0.2mm);

 \fill[black](3.0, 0)  circle (0.2mm);
\fill[black](3.055, 0.4)  circle (0.2mm);
\fill[black](3.055, -0.4)  circle (0.2mm);

\fill[black](3.428, 1.0)  circle (0.2mm);
\fill[black](3.671, 1.25)  circle (0.2mm);
\fill[black](3.962, 1.4)  circle (0.2mm);

\fill[black](3.428, -1.0)  circle (0.2mm);
\fill[black](3.671, -1.25)  circle (0.2mm);
\fill[black](3.962, -1.4)  circle (0.2mm);


\fill[black](7.5, 0)  circle (0.2mm);
\fill[black](7.555, 0.4)  circle (0.2mm);
\fill[black](7.555, -0.4)  circle (0.2mm);

\fill[black](7.928, 1.0)  circle (0.2mm);
\fill[black](8.171, 1.25)  circle (0.2mm);
\fill[black](8.462, 1.4)  circle (0.2mm);

\fill[black](7.928, -1.0)  circle (0.2mm);
\fill[black](8.171, -1.25)  circle (0.2mm);
\fill[black](8.462, -1.4)  circle (0.2mm);


\fill[black](12.0, 0)  circle (0.2mm);
\fill[black](12.055, 0.4)  circle (0.2mm);
\fill[black](12.055, -0.4)  circle (0.2mm);

\fill[black](12.428, 1.0)  circle (0.2mm);
\fill[black](12.671, 1.25)  circle (0.2mm);
\fill[black](12.962, 1.4)  circle (0.2mm);

\fill[black](12.428, -1.0)  circle (0.2mm);
\fill[black](12.671, -1.25)  circle (0.2mm);
\fill[black](12.962, -1.4)  circle (0.2mm);
\end{tikzpicture}
\caption{Visualisation of the relation \eqref{maintool}. The diagram on the left corresponds to the $N$-pt necklace block. One is represented as a sum of the $(N+2)$-pt comb channel blocks with two additional operators $\partial^m \cO_{\alpha}$. Red dots are depicted derivatives with respect to $z_0$ and $z_{N+1}$. }
\label{pic}
\end{figure}
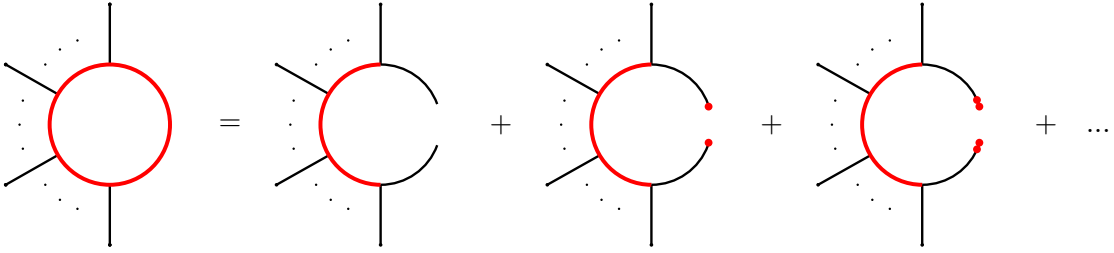

Fig.\bref{pic} illustrates the formula \eqref{maintool}. Notice that concerning the channel topology the necklace $N$-point block and the $(N+2)$-pt comb channel block differ only in the additional averaging over the conformal family of the operator $\cO_{\alpha}$ in the first line of \eqref{maintool}. It is implemented by considering the sum of comb channel blocks with descendant operators  $\partial^m \cO_{\alpha}$ in the first and last place. From the topology standpoint, the limits in the second line of \eqref{maintool} can be viewed by gluing endpoints in each comb block together. Finally, in the limit $q \rightarrow 0$ the necklace block reads
\be
\label{lq}
q^{-h_{\alpha}}\cF_{N}(q, z_1, ..., z_N) \Big|_{q \rightarrow 0}= \lim \limits_{\substack{%
\\
z_0 \to 0\\
z_{N+1} \to 0}} z_0^{-2 h_{\alpha}} \dps G_{N+2}(1/z_0, z_1, ...,z_{N}, z_{N+1}),
\ee
 which was  studied in details in \cite{Alkalaev:2022kal}.

On the sphere, global conformal blocks in different channels were studied earlier in the literature \cite{ Alkalaev:2017bzx, Kraus:2017ezw, Rosenhaus:2018zqn, Gerbershagen:2021yma}. It was found \cite{Rosenhaus:2018zqn} that the comb channel block can be expressed in terms of the so-called comb function defined by
\begin{multline}
     \label{CombDef1}
\FK{5}{3}{a_1\kk, b_1\kk,\!\ldots\kk, b_{k-1}\kk, a_2}{c_1\kk, \!\ldots\kk, c_k}{x_1\kk,\!\!\ldots\kk, \!x_k} \\
= \sum_{n_1, \ldots, n_k=0}^{\infty} \!\!\frac{(a_1)_{n_1} (b_1)_{n_1 + n_2} (b_2) _{n_2+n_3} \cdots (b_{k-1})_{n_{k-1} +n_k} (a_2)_{n_k}}{(c_1)_{n_1}\cdots (c_k)_{n_k}}\frac{x_1^{n_1}}{n_1!}\! \cdots\!\frac{x_k^{n_k}}{n_k!}~.
\end{multline}

For the set of dimensions we have chosen, the comb channel block $G_{N+2}$ in \eqref{maintool} has the form
\be
\label{Npt}
\ba{c}
G_{N+2} (z_0,\mathbf{z}, z_{N+1}|h, \tilde h, h_{\alpha}) = \dps \left(\frac{z_{12} z_{N-1,N}}{z_{01} z_{02} z_{N,N+1} z_{N-1,N+1}}\right)^{h_{\alpha}} \prod^{N-1}_{i=0}  \left(\frac{z_{i,i+2}}{z_{i,i+1}z_{i+1,i+2}} \right)^{h_{i+1}} \\
\\
\dps \times \prod^{N-2}_{i=0} \chi_{i}^{\tilde{h}_i} ~ \FK{5}{3}{-s_1\kk ~ b_0\kk  \ldots \kk b_{N-3} \kk ~  -s_2}{c_0 \kk~ \ldots \kk~ c_{N-2}}{\chi_0 \kk ~... \kk ~ \chi_{N-2}} ,  \quad \chi_{i} = \frac{z_{i,i+1} z_{i+2, i+3}}{z_{i, i+2} z_{i+1, i+3}},
\ea
\ee
where $F_{K}$ is \eqref{CombDef1}, $z_{i,j} = z_i - z_j$ and
\be
\label{a}
s_1 = h_1 - h_{\alpha} - \tilde{h}_0, \quad s_2 = h_{_N} - h_{\alpha} - \tilde{h}_{_{N-2}},
\ee
\be
\label{b}
b_i = \tilde h_{i} + \tilde h_{i+1} - h_{i+2}, \qquad i = 0,...,N-3,
\ee
\be
c_j = 2\tilde h_j , \qquad j = 0,..., N-2.
\ee
Notice that for future needs we have changed the sign in the notation for the first and last parameters of the comb function compared to the definition given in \cite{Rosenhaus:2018zqn}.

\section{Explicit block functions}
\label{sec:ebf}
In this section, we obtain necklace block functions using the formula \eqref{maintool}. Starting from the well-elaborated case of the $1$-pt block  \cite{Poghossian:2009mk, Hadasz:2009db,Kraus:2017ezw}, we analyze
particular examples of $2$- and $3$-pt blocks and generalize them to $N$-pt blocks.

\subsection{$1$-pt torus block}
The global $1$-pt torus block was initially considered in the context of large-$c$ limit of Virasoro conformal blocks \cite{Hadasz:2009db}.\footnote{From holographic perspective the torus blocks (as well as its $\mathcal{N} =1$ SUSY analogues) considered in \cite{Alkalaev:2016ptm, Alkalaev:2017bzx, RamosCabezas:2020mew, Alkalaev:2020yvq, Fortin:2020yjz, Fortin:2020zxw, Belavin:2022bib}. } Regarding the Casimir approach \cite{Kraus:2017ezw} it was shown that the $1$-pt torus block is subjected to the second order differential equation in $q$.

To calculate the $1$-pt torus block in accordance with \eqref{maintool}, we consider the $3$-pt block of primary operators with dimensions $(h_{\alpha}, h, h_{\alpha})$ inserted at points $(z_0, z_1, z_2)$ which has the following form
\be
\label{3f1}
G_{3} (z_0, z_1, z_{2}|h, h_{\alpha}) = (z_0 -z_1)^{-h} (z_1 - z_2)^{-h} (z_0 - z_2)^{h-2 h_{\alpha}}.
\ee
The torus matrix element in the second line of \eqref{maintool} was found to be \cite{Hadasz:2009db}
\be
A^{(3)}_m (z_1|h, h_{\alpha}) = z_1^{-h} \sum^{m}_{k=0} \frac{(m!)^2}{(k!)^2 (m-k)!} \frac{ (h-k)_{2k} (2h_{\alpha})_n}{(2h_{\alpha})_k},
\ee
so the first line in \eqref{maintool} yields
\be
\label{1pt}
\ba{c}
\cF_{1} (q, z_1|h, h_{\alpha}) = \dps \frac{  z_1^{-h} q^{h_{\alpha}}}{1-q} ~  _2F_1 \left(h, 1 - h , 2 h_{\alpha}, \frac{q}{q-1}\right) = \\
\\
\dps \frac{ z_1^{-h} q^{h_{\alpha}}}{(1-q)^{1-h}} ~  _2F_1 (h, 2h_{\alpha} +h -1, 2 h_{\alpha}, q) = \dps \frac{ z_1^{-h} q^{h_{\alpha}}}{(1-q)^{h}} ~  _2F_1 (1-h, 2h_{\alpha} -h, 2 h_{\alpha}, q).
\ea
\ee
Here the formulas in the last line are obtained by applying the Pfaff transformations \cite{zbMATH03219699} to the hypergeometric function in the first line. Notice that for $h = 2 h_{\alpha}$ the block \eqref{1pt} reduces to
\be
\label{1ptp}
\cF_{1} (q, z_1|2 h_{\alpha}, h_{\alpha}) =  q^{h_{\alpha}} (1-q)^{-2h_{\alpha}} z^{-2 h_{\alpha}}_1,
\ee
which is going to be useful for the further consideration.

\subsection{$N$-pt necklace block - preliminaries and assumptions}

In order to approach $N$-pt necklace blocks it is important to mentioned that the key role in computing via \eqref{maintool} is played by the dependence on variables $z_0$ and $z_{_{N+1}}$ in the comb channel block. For the comb function we can separate $z_0$ and $z_{_{N+1}}$ with the help of
splitting identities \cite{Rosenhaus:2018zqn}
\be
\label{combforfuture}
\ba{c}
\dps \FK{5}{3}{-s_1\kk ~ b_0\kk  \ldots \kk b_{N-3} \kk ~  -s_2}{c_0 \kk~ \ldots \kk~ c_{N-2}}{\chi_0 \kk ~... \kk ~ \chi_{N-2}} = \sum^{\infty}_{n_0, n_{N-2}=0} \frac{(-s_1)_{n_{_0}}  (b_{0})_{n_{_0}} (b_{N-3})_{n_{N-2}} (-s_2)_{n_{N-2}} }{n_0 ! n_{N-2}! (c_0)_{n_0} (c_{_{N-2}})_{n_{_{N-2}}}}
\\
\\
\dps \times \chi^{n_0}_0 \chi^{n_{_{N-2}}}_{_{N-2}} \FK{5}{3}{ b_0+ n_0\kk ~ b_1\kk~   \ldots \kk b_{N-4} \kk ~  b_{N-3} + n_{_{N-2}} }{c_1 \kk~ \ldots \kk~ c_{N-3}}{\chi_1 \kk ~... \kk ~ \chi_{N-3}},
\ea
\ee
so for the $(N+2)$-pt comb block \eqref{Npt} one has
\be
\label{assum}
\ba{c}
\dps G_{N+2} (z_0,\mathbf{z}, z_{N+1}|h, \tilde h, h_{\alpha}) \sim  z^{-s_1 - 2h_{\alpha}}_{_{01}} z_{_{02}}^{s_1} \; z^{s_2}_{_{N-1,N+1}} z_{_{N, N+1}}^{-s_2 - 2h_{\alpha}}
\\
\\
\dps \times \sum^{\infty}_{n_0, n_{N-2}=0} \frac{(-s_1)_{n_{_0}}  (b_{0})_{n_{_0}} (b_{N-3})_{n_{N-2}} (-s_2)_{n_{N-2}} }{ n_{_0} ! n_{_{N-2}}!
 (c_0)_{n_0} (c_{_{N-2}})_{n_{_{N-2}}}} \chi^{n_0}_0 \chi^{n_{_{N-2}}}_{_{N-2}}, \quad  \chi_0 \sim \frac{z_{01}}{z_{02}}, ~ \chi_{_{N-2}} \sim \frac{z_{_{N,N+1}}}{z_{_{N-1,N+1}}}.
\ea
\ee
One can see that direct implementation of \eqref{maintool} to the formula above gives complex expressions. In order to simplify calculations and obtain explicit block functions, we consider particular cases
\be
\label{mi}
 s_{1,2} = 0, 1, 2, ...,
\ee
where $s_{1,2}$ are given by \eqref{a}. Under this assumption, the double sum in the second line of \eqref{assum} is reduced to the product of finite ones, which contain $(s_1+1)$ and $(s_2+1)$ terms, respectively. Notice  a condition $b_{N-3} = - t_1, b_{0} = - t_2$, $t_{1,2} = 0, 1, 2, ...$ can be considered by the same reasons. Despite this, we will still concentrate on the condition \eqref{mi}, because in the simplest case $t_{1,2} =0$ we arrive at $G_{N+2} \sim z^{-2 h_{\alpha}}_{01} z^{-2h_{\alpha}}_{N,N+1}$, while for $b_{N-3} = b_{0} = 0$ one has a more complex structure due to $s_{1,2} \neq 0$. After setting \eqref{assum}, $G_{N+2}$ reduces to a  polynomial (multiplied by the factor $z^{-2h_{\alpha}}_{_{01}} z_{_{N, N+1}}^{- 2h_{\alpha }} $) in the variables $z_{02}/z_{01}$ and $z_{N-1, N+1}/z_{N, N+1}$.

It is important to compare conformal blocks under conditions \eqref{mi} and ones with singular intermediate operators $\tilde{h} = - j/2, \; j = 0, 1, 2,..$. These operators have finite-dimensional conformal families leading to finite sums in expressions for conformal blocks which seems similar to \eqref{assum}. However, these cases are distinct - we see that in general the condition \eqref{mi} 
 hold for non-singular operators. Moreover, conformal blocks with singular operators realise a particular case of \eqref{mi} where one operator has a negative integer dimension and two other have half-integers,  or all three operators have negative integer dimensions. 

The computation of the $N$-pt necklace block under the assumption \eqref{mi} will be illustrated with concrete examples in the next section. We  consider obtaining of $2$- and $3$-pt blocks separately, since the application of \eqref{mi} to the corresponding comb blocks $G_{N+2}$ gives power functions. In what follows, we slightly change the notation for the necklace block on the torus and move $(s_1, s_2)$ to the superscript
\be
\label{red}
\cF^{(s_1, s_2)}_{N} (q, \mathbf{z}|h, \tilde h, h_{\alpha})  \equiv \cF_{N} (q, \mathbf{z}|h, \tilde h, h_{\alpha}), \qquad \text{if} ~~~ s_{1,2} =0, 1, 2,...
\ee

\subsection{2-pt and 3-pt necklace blocks}
\label{sec:ttr}
\paragraph{2-pt block.}
The  $4$-pt comb channel block \cite{Dolan:2003hv, Dolan:2011dv, SimmonsDuffin:2012uy} associated with a $2$-pt torus block in the necklace channel reads
\be
\label{4pt_pl}
\ba{l}
\dps
G_{4} (z_0,z_1, z_2, z_3| h, \tilde h, h_{\alpha}) =z_{01}^{-h_{\alpha}-h_1}z_{23}^{-h_2 - h_{\alpha}} z_{02}^{-h_{\alpha}+h_{1}}z_{12}^{2 h_{\alpha}-h_{1}-h_{2}} z_{13}^{h_{2}-h_{\alpha}}
\vspace{3mm}
\\
\dps
\hspace{10mm}\times\,  \chi_0^{\tilde h_0}  \;{}_2F_1 \left( \tilde h_0 + h_{\alpha} - h_1, \tilde h_0 + h_{\alpha} -h_2, 2 \tilde h_0\,;\, \chi_0 \right), \qquad \chi_0 = \frac{z_{01} z_{23}}{z_{02} z_{13}}\;.
\ea
\ee
 The block $G_{4} (z_0,..., z_{3}| h, \tilde h, h_{\alpha})$ is parameterized  by 4 external dimensions $(h_{\alpha}, h_1, h_2, h_{\alpha})$ and one intermediate dimension $\tilde h_0$ which are denoted by a set $ (h, \tilde h, h_{\alpha})$. Notice that in this case, the formula \eqref{combforfuture} is not applicable explicitly since we only have one summation from a hypergeometric function ${}_2F_1$ in  \eqref{4pt_pl}. Despite this, the general scheme remains the same - setting the first (or the second) parameter of the hypergeometric function to be 0 or a negative integer, one obtains a  polynomial instead of an infinite sum
\be
{}_2F_1 \left( -s, b, c \,;\, x \right) = \sum^{s}_{i=0} \,  (-1)^i \, C^{i}_{s} \, \frac{(b)_i}{(c)_i} \, x^i.
\ee
In addition to \eqref{mi} for further simplification we set\footnote{Without such a requirement the comb channel block \eqref{4pt_pl} has a more complex form. Assuming $\tilde h_0 + h_{\alpha} - h_1 = -s_1, ~ \tilde h_0 + h_{\alpha} - h_2 = -s_2$, $s_{1,2} = 0, 1, 2,...$ yields
\be
G_{4} (z_0,z_1, z_2, z_3| h, \tilde h, h_{\alpha}) \sim  z_{01}^{-2 h_{\alpha} -s_1 } ~z^{s_1}_{02} ~  z_{12}^{2 h_{\alpha}-2 h_{1}+ s_1 - s_2} ~   z_{13}^{s_2} ~ z_{23}^{-s_2 - 2 h_{\alpha}}.
\ee}
\be
 h_1 = h_2 \equiv h ~~ \rightarrow ~~ s_1 = s_2 \equiv s,
\ee
which results in
\be
\label{g4m}
G_{4} (z_0,...,z_3| h, \tilde h, h_{\alpha})= z^{-s-2 h_{\alpha}}_{01} z^{s}_{02} ~  z_{12}^{2 h_{\alpha} - 2 h} z^s_{13} ~ z^{-s - 2 h_{\alpha}}_{23} \sum^{s}_{i=0} \,  \frac{i!
(C^{i}_{s})^2}{(2 h_{\alpha})_i} \, \left( \frac{z_{01} z_{23}}{z_{02} z_{13}}\right)^i.
\ee
The simplest case corresponds to $s=0$  where \eqref{g4m} reduces to
\be
G_{4} (z_0,..., z_3| h, h_{\alpha})\Big|_{s=0} = z^{-2h_{\alpha}}_{01}  z^{2h_{\alpha}-2h}_{12} z^{-2h_{\alpha}}_{23},
\ee
so the second line of \eqref{maintool} gives
\be
A^{(4)}_m (z_1, z_2|h, h_{\alpha}) = \left((2 h_{\alpha})\right)^2 z_2 ^{-2 h_{\alpha}}\left(\frac{z_1}{z_2}\right)^{k} z_{12}^{2 h_{\alpha}-2 h}.
\ee
Here, $\tilde h_0$-dependence is excluded by substituting of the condition $s=0$. Applying the first line of \eqref{maintool} to the formula above we find the $2$-pt necklace block
\be
\label{t0}
\cF^{(0)}_{2}(q, z_1, z_2|h, h_{\alpha}) = z_{12}^{2 h_{\alpha}-2 h} q^{h_{\alpha}} \left(1- \dps \frac{q z_1}{z_2}\right)^{-2 h_{\alpha}} z_2^{-2 h_{\alpha}}.
\ee
Analogously, the result for the case $s = 1$ reads
\be
\ba{c}
\label{t1}
\dps \cF^{(1)}_{2}(q, z|h,\tilde h_0, h_{\alpha}) =  \frac{\cF^{(0)}_{2}(q, z|h, h_{\alpha})\left(1- \dps q x\right)^{-2}}{2 \tilde h_0 h_{\alpha}}
\\
\\
\dps \times \left[ h_{\alpha} q^2  x (2 \tilde h_0 + x)+q (\tilde h_0 x (x -2 -4 h_{\alpha})+\tilde h_{0}-2 h_{\alpha} x)+2 \tilde h_{0} h_{\alpha} x+h_{\alpha} \right], \quad x \equiv \frac{z_1}{z_2}.
\ea
\ee
There are several comments to make about these formulas. First, one can see that in \eqref{t0} the first factor is the $2$-pt correlation function of two primary operators with conformal dimensions $h-h_{\alpha}$. In Section \bref{sec:CE}  we prove that this property generalizes to the $N$-pt necklace block on the torus. Second, the $2$-pt block \eqref{t1} factorizes into the product of the $s=0$ block \eqref{t0} and a polynomial in the variables $q, x$. Moreover, for cases  $s = 2, 3$ (see Appendix \bref{sec:ma2} for details) one can see that $\cF^{(s)}_{2}$ has the following structure
\be
\label{obs1}
\dps \cF^{(s)}_{2}(q, z|h,\tilde h_0, h_{\alpha}) = \dps \cF^{(0)}_{2}(q, z|h,  h_{\alpha}) (1- q x)^{-2s} \tilde{P}^{(s)}(q, x),
\ee
where $\tilde{P}^{(s)}(q, x)$ is a polynomial of degree at most $2s$ in $q$ and $x$. Finally, for polynomials $\tilde{P}^{(s)}(q, x)$ one can show that
\be
\label{POLPR}
 \tilde{P}^{(s)}(q, x) = (-q x)^{2s} \tilde{P}^{(s)}(1/q, 1/x), 
\ee
which provides that the bare block $\cF^{(s)}_{2}(q, \mathbf{z})$ is invariant under the change 
\be
\label{tna}
x \to x^{-1} \; \; \text{or} \;\; z_1 \leftrightarrow z_2, \; \qquad q \to q^{-1}\;. 
\ee
This symmetry can be understood as follows: we see that the second transformation \eqref{tna} maps an annulus with boundaries $|q|$  and $1$ to one with boundaries at $1$ and $|q|^{-1}$. Notice that this map itself provides a swap of $z_1$ and $z_2$. These two torus appear to be two choices of the factor-space construction resulting in the same manifold. Hence, the symmetry \eqref{tna} emerges as a consequence of the identification producing the torus from $\mathbb{R}^2$. By the same reasoning, the similar symmetry (involving several $x_i \rightarrow 1/x_i$ transformations) exists for $N>2$.

\paragraph{3-pt block.}
In this case the $5$-pt comb channel block \eqref{Npt} is given by
\be
\ba{c}
G_{5} (z_0, ...,z_4|h, \tilde h, h_{\alpha}) = z^{-s_1 - 2 h_{\alpha}}_{01} z^{s_1}_{02} z^{h_{\alpha} + \tilde h_1 -h_2 - h_3}_{12} z^{h_3 - \tilde h_0 - \tilde h_1}_{13} z^{\tilde h_0-h_2-\tilde h_1}_{23} z^{s_2}_{24} z^{-s_2 - 2 h_{\alpha}}_{34} \\
\\
\dps \times \FK{5}{3}{-s_1\kk ~ \tilde h_0 + \tilde h_1 - h_2 \kk ~  -s_2}{2 \tilde h_0 \kk~ 2\tilde h_{1}}{\chi_0 \kk ~ \chi_{1}},
\ea
\ee
where $s_1$ and $s_2$ are defined by \eqref{a} and $F_{K}$ is the Appell function $F_2$ \cite{Rosenhaus:2018zqn, zbMATH03219699}. As previously, it can be presented as a finite sum if $s_{1,2}$ are positive integers
\be
\FK{5}{3}{-s_1\kk ~ \tilde h_0 + \tilde h_1 - h_2 \kk ~  -s_2}{2 \tilde h_0 \kk~ 2\tilde h_{1}}{\chi_0 \kk ~ \chi_{1}} = \sum^{s_1, s_2}_{n_0, n_1 =0} (-1)^{n_0 + n_1} C^{n_0}_{s_1} C^{n_1}_{s_2} \frac{(\tilde h_0 + \tilde h_1 - h_2)_{n_0 +n_1}}{(2\tilde h_0)_{n_0} (2\tilde h_1)_{n_1}} \chi^{n_0}_0 \chi^{n_1}_1.
\ee
In contrast to the previous case, the condition $s_1 \neq s_2$ does not lead to complications so $s_1$ and $s_2$ can be independent. For $s_1 = s_2 =0$ the double sum above reduces to 1 and using \eqref{maintool} we have
\be
\ba{c}
\label{tr0}
\cF^{(0,0)}_{3}(q, z_1, z_2, z_3|h, \tilde{h}, h_{\alpha}) = q^{h_{\alpha}} \left(1-\dps \frac{q z_1}{z_3}\right)^{-2 h_{\alpha}}  z_3^{-2 h_{\alpha}} \tilde{G}_3 (z_1, z_2, z_3| h, \tilde h), \\
\\
\tilde{G}_3 (z_1, z_2, z_3| h, h_{\alpha}) = z_{12}^{h_3-h_2- h_1} z_{13}^{h_2-h_1- h_3+2 h_{\alpha}} z_{23}^{h_1 - h_2 - h_3}.
\ea
\ee
Few comments are in order. First, one can see that the structure of the first factor resembles one in \eqref{t0} with $z_2$ replaced by $z_3$.  Second, $\tilde{G}_3$ is nothing more than a plane $3$-pt conformal block of primary operators with dimensions $(h_1 - h_{\alpha}, h_2,  h_3 - h_{\alpha})$, located at points $(z_1, z_2, z_3)$. For $s_{1,2} \neq 0$ the $3$-pt necklace blocks and their  properties are listed in Appendix \bref{sec:ma3}.

\subsection{$N$-pt necklace block}
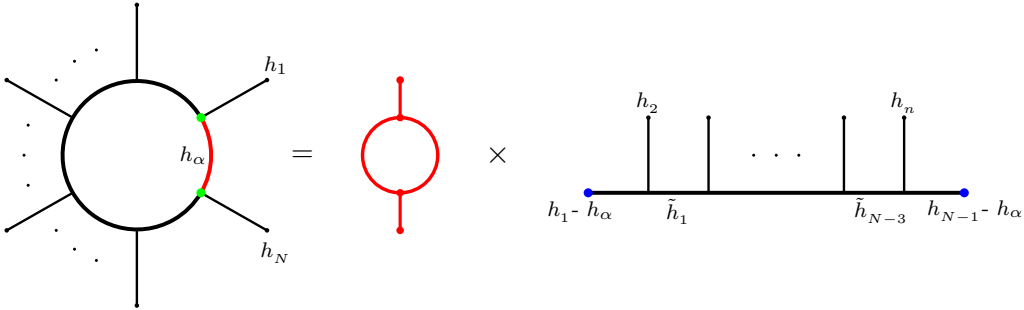
\begin{figure}[H]
\centering
\begin{tikzpicture}[scale=1.0]
 \tkzDefPoint(0,0){OO}
  \tkzDefPoint(1.0,0){B}
    \tkzDefPoint(2,0){A}
      \tkzDefPoint(0.8515, -0.5){C}


 \draw[color = black, line width  = 0.9] (0, -1) -- (0, -2.0);
 \draw[color = black, line width  = 0.9] (0, 1) -- (0, 2.0);

  \draw[color = black, line width  = 0.9] (-0.8515, -0.5) -- (-1.73, -1.0);
  \draw[color = black, line width  = 0.9] (-0.8515, 0.5) -- (-1.73, 1.0);

  \tkzDrawArc[rotate,color=black,line width  = 1.5](OO,C)(-300)
\tkzDrawArc[rotate,color=red,line width  = 1.5](OO,C)(60)

    \draw[color = black, line width  = 0.9] (0.8515, -0.5) -- (1.73, -1.0);
  \draw[color = black, line width  = 0.9] (0.8515, 0.5) -- (1.73, 1.0);

\fill[green] (0.8515, -0.5) circle (0.6mm);
\fill[green] (0.8515, 0.5) circle (0.6mm);

\fill[black](-1.73, -1.0) circle (0.3mm);
\fill[black](-1.73, 1.0)circle (0.3mm);
\fill[black](1.73, 1.0)circle (0.3mm);
\fill[black](1.73, -1.0) circle (0.3mm);
\fill[black](0, -2) circle (0.3mm);
\fill[black](0, 2)  circle (0.3mm);

\fill[black](-1.5, 0)  circle (0.2mm);
\fill[black](-1.445, 0.4)  circle (0.2mm);
\fill[black](-1.445, -0.4)  circle (0.2mm);

\fill[black](-1.072, 1.0)  circle (0.2mm);
\fill[black](-0.829, 1.25)  circle (0.2mm);
\fill[black](-0.538, 1.4)  circle (0.2mm);

\fill[black](-1.072, -1.0)  circle (0.2mm);
\fill[black](-0.829, -1.25)  circle (0.2mm);
\fill[black](-0.538, -1.4)  circle (0.2mm);

  \draw  (2.2, 0.0) node {$=$};

 \draw  (0.7, 0.0) node {\scriptsize{ $h_{\alpha}$}};

  \draw  (1.8, 1.2) node {\scriptsize{ $h_{1}$}};
  \draw  (1.8, -1.3) node {\scriptsize{ $h_{_N}$}};

\tkzDefPoint(3.5,0){BB}
\tkzDefPoint(4.0,0){AA}

\tkzDrawCircle[red, line width  = 1.3](BB,AA)

 \draw  (4.8, 0.0) node {\large $\times$};
\draw [red, line width=1.2] (3.5,-0.5) -- (3.5,-1.0);
\draw [red, line width=1.2] (3.5,0.5) -- (3.5,1.0);

\draw [black, line width=1.5] (6,-0.5) -- (11,-0.5);

\fill[blue] (6, -0.5) circle (0.6mm);
\fill[blue] (11, -0.5) circle (0.6mm);

\draw [black, line width=0.9] (6.8,-0.5) -- (6.8,0.5);
\draw [black, line width=0.9] (7.6,-0.5) -- (7.6,0.5);

\draw [black, line width=0.9] (9.4,-0.5) -- (9.4,0.5);
\draw [black, line width=0.9] (10.2,-0.5) -- (10.2,0.5);

\fill[red] (3.5,0.5)  circle (0.5mm);
\fill[red] (3.5,-0.5)  circle (0.5mm);
\fill[red] (3.5,1)  circle (0.5mm);
\fill[red] (3.5,-1)  circle (0.5mm);

\fill[black] (6.8,0.5)  circle (0.3mm);
\fill[black] (7.6,0.5)  circle (0.3mm);
\fill[black] (9.4,0.5) circle (0.3mm);
\fill[black] (10.2,0.5) circle (0.3mm);

\fill[black] (8.2,0)  circle (0.2mm);
\fill[black] (8.5,0)  circle (0.2mm);
\fill[black] (8.8,0) circle (0.2mm);

 \draw  (6.8,0.7)  node {\scriptsize$h_{_2}$};
 \draw  (10.2,0.7)  node {\scriptsize$h_{_n}$};

  \draw  (5.9, -0.75)  node {\scriptsize$h_{_1}$-~$h{_{\alpha}}$};
  \draw  (11.15, -0.75)  node {\scriptsize$h_{_{N-1}}$-~$h{_{\alpha}}$};

    \draw  (7.2, -0.75)  node {\scriptsize$\tilde{h}_{_1}$};
    \draw  (9.9, -0.75)  node {\scriptsize$\tilde{h}_{_{N-3}}$};

\end{tikzpicture}
\caption{The $N$-pt necklace channel block under conditions $ h_{\alpha} + \tilde{h}_0 - h_1 = h_{\alpha} + \tilde{h}_{_{N-2}} - h_{_N} =0, $ which are shown by the green dots in the left picture. According to the \eqref{SNpt}, this block factorizes into the product of the function \eqref{Pfunction} (red) and the $N$-pt comb block with specific dimensions depicted by blue dots.}
\label{f2}
\end{figure}

For $s_1 = s_2 = 0$ the computation of the $N$-pt necklace block mimics one for the $3$-pt block described above. After applying \eqref{maintool} to \eqref{Npt}, the $N$-pt necklace block is found to be
\be
\label{SNpt}
F^{(0, 0)}_{N}(q, \mathbf{z}|h, \tilde h, h_{\alpha}) = P^{(0)}(q, z_1, z_N|h_{\alpha}) \tilde{G}_{N} (\mathbf{z}|h, \tilde h),
\ee
where
\be
\label{Pfunction}
P^{(0)}(q, z_1, z_{_N}|h_{\alpha}) = q^{h_{\alpha}} \left(1-\frac{q z_{_1}}{z_{_N}}\right)^{-2 h_{\alpha}}  z^{-2 h_{\alpha}}_{_N},
\ee
and $\tilde{G}_{N} (\mathbf{z}|h, \tilde h)$ is the particular $N$-point comb channel block with a set of external dimensions $(h_1-h_{\alpha}, ..., h_N - h_{\alpha})$ and $(N-3)$ intermediate dimensions $(\tilde h_1,...,\tilde h_{N-3})$ (see Fig. \bref{f2}).

The structure of the block \eqref{SNpt} remotely resembles one for the $N$-pt OPE channel block on the torus which was discussed in \cite{Kraus:2017ezw}. There, the $N$-pt OPE block (for any set of conformal dimensions) is factorized into a product of  the $1$-pt torus block \eqref{1pt} and the $(N+2)$-pt plane block in some channel which was determined by the OPE's topology of the torus block and could not be a comb channel (see section 4.1. in \cite{Kraus:2017ezw} for details). For the necklace channel blocks on the torus, we have the function \eqref{Pfunction} instead of the $1$-pt block and the second factor is the $N$-pt block in the comb channel with the first and the last dimensions shifted by $h_{\alpha}$. It is also interesting that the 1-pt torus block \eqref{1ptp} obtained by applying the condition $h=2h_{\alpha}$ (which is analogous to \eqref{mi} coincides with \eqref{Pfunction} for $z_1 = z_N$. Finally, we see a branch cut in terms of the composite variable $q z_1/z_N$. 

For $s_{1,2} \neq 0$ the $N$-pt necklace block \eqref{maintool} can be cast into the following form
\be
\ba{c}
\label{generalNpt}
\cF^{(s_1, s_2)}_{N}(q,\mathbf{z}|h, \tilde h, h_{\alpha})  = q^{h_{\alpha}} \dps \sum^{s_1, s_2}_{k, l =0} \frac{(-s_1)_k  (b_{0})_k (b_{N-3})_{l} (-s_2)_{l} } {k! l! (c_0)_{k} (c_{_{N-2}})_{l}} \tilde{G}^{(k,l)}_N (\mathbf{z}| h, \tilde h)  B^{(k,l)} (q, \mathbf{z}), \\
\\
B^{(k,l)} (q, \mathbf{z}) = z^{k-s_1}_{12} z^{ l- s_2 }_{N-1, N} \left(\frac{z_{_{N-1}}}{z_{_N}}\right)^{s_2 -l}  z^{-2 h_{\alpha}}_{_N} \dps \sum^{\infty}_{m=0} \dps \frac{(2 h_{\alpha}+s_2 -l)_m }{m!(2 h_{\alpha})_m}  \left(\frac{q z_1}{z_{N}}\right)^m    \times
\\
\\
 _2F_1\left(l-s_2,-m, 1 -2 h_{\alpha}+l-m-s_2;\dps \frac{z_N}{z_{N-1}}\right) \dps \sum^{m}_{p=0} C^{p}_m (2 h_{\alpha} +s_1 -k)_{m-p} (k - s_1)_{p} \left(\frac{z_2}{z_1}\right)^p,
\ea
\ee
where $\tilde{G}^{(k,l)}_N (\mathbf{z}| h, \tilde h)$ stands for the $N$-pt comb channel block with following external  $(h_1-h_{\alpha}-(s_1 -k),...,h_N-h_{\alpha}-(s_2 -l))$ and intermediate $(\tilde h_1, ..., \tilde h_{N-3})$ dimensions. Note that if $s_{1,2} =0$ which implies $k=0=l$ we get
\eqref{Pfunction} from the second and third line in \eqref{generalNpt}. For $s_{1,2} \neq 0$  the necklace block is structurally a finite sum of $N$-pt comb blocks with the first and last dimensions shifted.

The formula \eqref{generalNpt} provides an efficient tool for calculating necklace blocks for predefined $s_{1,2}$. Although we have not been able to find closed expressions for arbitrary integers $s_{1,2}$, the explicit functions \eqref{generalNpt} are useful as the first examples of $N$-pt necklace block functions obtained. It also can shed light on the structure of the $N$-pt block so with the right choice of variables, the block may reduce to known special functions.

In the OPE limit $z_{_N} \rightarrow z_{_{N-1}}$, the behaviour of the necklace block coincides with one for the comb channel block. Indeed, the prefactors $B^{(k,l)}$ behave as $z^{ l- s_2 }_{N-1, N}$ at $z_{_N} \rightarrow z_{_{N-1}}$, so in the leading order for the $N$-pt necklace block we have
\be
\label{OL}
\cF^{(s_1, s_2)}_{N}(q, z_1, ..., z_{_N}) \rightarrow z^{\tilde h_{_{N-3}} - h_{_N} - h_{_{N-1}}}_{_{N-1,N}} \cF^{(s_1, s_2)}_{N-1}(q, z_1, ..., z_{_{N-1}}) , \quad z_{_N} \rightarrow z_{_{N-1}},
\ee
where one uses that $\tilde{G}^{(k,l)}_N (z_1, ..., z_{N}) \rightarrow z^{\tilde h_{_{N-3}} - h_{_N} - h_{_{N-1}} + s_2 - l}_{_{N-1,N}} \tilde{G}^{(k,l)}_{N-1} (z_1, ..., z_{N-1})$.

\section{Casimir equations}
\label{sec:CE}
The $N$-pt necklace block $\cF_{N}(q,\mathbf{z}|h, \tilde h, h_{\alpha})$ is subjected by the Casimir equations \cite{Alkalaev:2022kal}
\be
\label{CAS}
\ba{l}	
~~~~~\dps \bigg[q^2\partial_q^2 - \frac{2q}{1-q}q\partial_q +\frac{q}{(1-q)^2}\sum_{i=1}^N\cL_{-1}^{(i)}\sum_{k=1}^N\mathcal{L}_{1}^{(k)}  - h_{\alpha} ( h_{\alpha}-1) \bigg] \cF_{N}(q,\mathbf{z}|h, \tilde h, h_{\alpha}) = 0 ,
\vspace{4mm}
\\
\dps
\bigg[q^2\partial_q^2 - \frac{2q}{1-q}q\partial_q + \frac{1}{(1-q)^2}\bigg(\sum_{k=1}^{j-1}\mathcal{L}_{-1}^{(k)}+q\sum_{k=j}^{N}\mathcal{L}_{-1}^{(k)} \bigg)\bigg(q \sum_{k=1}^{j-1}\mathcal{L}_{1}^{(k)}+\sum_{k=j}^{N}\mathcal{L}_{1}^{(k)}\bigg) + \sum_{k=j}^N\mathcal{L}_{0}^{(k)}\sum_{l=j}^N\mathcal{L}_{0}^{(l)}
\vspace{2mm}
\\
\dps
~~+\left(2 q \partial_q - \frac{1+q}{1-q} \right) \sum_{k=j}^N\mathcal{L}_{0}^{(k)} - \tilde h_{j-2} (\tilde h_{j-2}-1) \bigg] \cF_{N}(q,\mathbf{z}|h, \tilde h, h_{\alpha}) = 0, \quad j=2,...,N,
\ea
\ee
which are supplemented by the Ward identity
\be
\label{Ward}
\sum^{N}_{i=1}\mathcal{L}^{(i)}_{0} \cF_{N}(q,\mathbf{z}|h, \tilde h, h_{\alpha}) = 0.
\ee
Here, the differential operators   $\mathcal{L}^{(i)}_{I}$ (in terms of $z_i$) are given by \eqref{DO}. Hence, the equations \eqref{CAS} and \eqref{Ward} together are combined into a system of $(N+1)$ partial differential equations for the function $\cF_{N}(q,\mathbf{z}|h, \tilde h, h_{\alpha})$.  The asymptotic behavior in $q$ at $q\rightarrow0$ is fixed by the definition \eqref{maintool}.\footnote{The second branch behaves as $q^{1-h_{\alpha}}$ at $q\rightarrow0$ which can be found from the first equation of \eqref{CAS} and correspond to the shadow intermediate operator with a dimension $1 - h_{\alpha}$.} One can reduce these equation to the $(N-1)$-pt necklace block using \eqref{OL} and see that they have the same form with change $N \rightarrow N-1$.  Next, we show that the necklace block functions computed in the previous section satisfy the Casimir equations \eqref{CAS}.

\paragraph{Lower-point blocks.}
In the context of the Casimir equations the $1$-pt block has been studied before \cite{Kraus:2017ezw}. For such a block we only have the first equation \eqref{CAS}  which can be reduced to a hypergeometric equation. The solution to this equation corresponding to the behavior $q^{h_{\alpha}}$ at $ q \rightarrow 0$ is given by \eqref{1pt}.  The factor $z_1^{-h}$ in \eqref{1pt} ensures that the given block satisfies the Ward identity $\mathcal{L}^{(1)}_0 \cF_{1} (q, z_1|h, h_{\alpha}) =0$.

Compared to the $1$-pt block, general $2$- and $3$-pt blocks are non-trivial functions of coordinates $z_i$ which can be mixed with $q$. Concerning the Ward identity \eqref{Ward} it is satisfied for the $2$-pt block \eqref{t0} and hence, it also holds for \eqref{t1} and for cases listed in Appendix \bref{sec:ma2} since they only depend on the ratio $x = z_1/z_2$. The same applies to the $3$-pt blocks found in Section \bref{sec:ttr} and Appendix \bref{sec:ma3}. Since the Casimir equations can be easily written out for $N=2,3$ blocks we straightforwardly check them using computer algebra. It also allows us to find and verify a number of properties that may be useful later (see also Appendices \bref{sec:ma2} and \bref{sec:ma3}). In particular, for $N=2$ the Casimir system \eqref{CAS} is invariant under the change $z_1 \leftrightarrow z_2, ~ q \rightarrow 1/q$ in the $2$-pt block function.

\paragraph{The $N$-pt necklace block.}
\label{sec:NPT}
Since the necklace block $\cF^{(s_1,s_2)}_{N} (q, \mathbf{z}|h, \tilde h)$ is expressed as a sum over comb channel blocks multiplied by functions $B^{(k,l)}$, one can exploit the fact that such comb blocks are determined by the system of (plane) Casimir equations. For the case $s_1 = s_2 =0$ we trace how the system \eqref{CAS} is reduced to Casimir equations for the comb block $\tilde{G}_{N} (\mathbf{z}|h, \tilde h)$. We also discuss the case $s_{1,2} \neq 0$ at the end of the section.

To start we remind that the comb channel block $\tilde{G}_{N} (\mathbf{z}|h, \tilde h)$ satisfies Ward's identities
\be
\ba{c}
\label{PWard}
\dps \sum^{N}_{i=1}\mathcal{L}^{(i)}_{-1} \tilde{G}_{N} (\mathbf{z}|h, \tilde h)  = 0, \quad \left (\sum^{N}_{i=1}\mathcal{L}^{(i)}_{0} - 2 h_{\alpha} \right) \tilde{G}_{N} (\mathbf{z}|h, \tilde h) = 0, \\
\\
\dps \left (\sum^{N}_{i=1}\mathcal{L}^{(i)}_{1} - 2 h_{\alpha} (z_1 + z_{_N})\right) \tilde{G}_{N} (\mathbf{z}|h, \tilde h)  =0,
\ea
\ee
where the extra terms in the last two equalities appear from shifting the conformal dimensions in the block (see discussion under \eqref{SNpt}). It is worth emphasizing that structures of this type but with a part of the operators are multiplied by $q$, are explicitly exuded in the Casimir equations \eqref{CAS}. Having \eqref{PWard}, one can easily prove the Ward identity \eqref{Ward} for the block \eqref{SNpt}
\be
\sum^{N}_{i=1}\mathcal{L}^{(i)}_{0} \cF^{(0,0)}_{N} (q, \mathbf{z}|h, \tilde h, h_{\alpha}) = \tilde G_{N} (\mathbf{z} |h, \tilde h) \left(z_1 \partial_1 + z_{_N} \partial_{_N} + 2 h_{\alpha} \right) P^{(0)}(q, z_1, z_{_N}|h_{\alpha}) = 0,
\ee
where the last equality is verified explicitly using \eqref{Pfunction}. Then, there are the following properties of the function $ P^{(0)}$ which resemble Ward identities  
\be
\ba{c}
\label{PR}
\left(\partial_1 + q \partial_{_N}\right) P^{(0)}(q, z_1, z_{_N}|h_{\alpha}) = 0, \quad \left(z_1 \partial_1 + z_{_N} \partial_{_N} + 2 h_{\alpha} \right) P^{(0)}(q, z_1, z_{_N}|h_{\alpha}) =0, \\
\\
\left( q (z^2_1 \partial_1 + 2 h_{\alpha} z_1 )+ z^2_{_N} \partial_{_N} + 2 h_{\alpha} z_{_N} \right) P^{(0)}(q, z_1, z_{_N}|h_{\alpha}) = 0.
\ea
\ee
Besides, one can find relations between derivatives with respect to $q$ and $z$
\be
\label{RP2}
\left(z_1 \partial_1 + h_{\alpha} + q \partial_q\right) P^{(0)}(q, z_1, z_{_N}|h_{\alpha}) = 0, \quad \left(z_{_N} \partial_{_N} + h_{\alpha} - q \partial_q\right) P^{(0)}(q, z_1, z_{_N}|h_{\alpha}) = 0. 
\ee

Returning to the equations \eqref{CAS} it is necessary to discuss how exactly they reduce to the Casimir equations for the $N$-pt comb channel block. For the first equation \eqref{CAS} the substitution of \eqref{PWard} gives us
\be
\ba{c}
\label{SP}
\dps \bigg[-q^2\partial_q^2 + \frac{2q}{1-q}q\partial_q + h_{\alpha} ( h_{\alpha}-1)  -\frac{q}{(1-q)^2}  \\
\\
\dps \times \left( ( \partial_1 + \partial_N ) (z^2_1\partial_1 + z^2_{N} \partial_N) + 2 h_{\alpha} (2 + (z_1+z_N) (\partial_1 + \partial_N))\right) \bigg]  P^{(0)}(q, z_1, z_{_N}|h_{\alpha}) = 0.
\ea
\ee
which is satisfied for $P^{(0)}(q, z_1, z_{_N}|h_{\alpha})$ defined by \eqref{Pfunction}. The second and the last ($j=N$) equations in \eqref{CAS}  after a little bit of algebra involving  \eqref{PWard}, \eqref{PR} and \eqref{RP2} can be cast into the form
\be
\ba{c}
\left( \mathcal{L}^{(1)}_1 \mathcal{L}^{(1)}_{-1} + \mathcal{L}^{(1)}_0 - \left( \mathcal{L}^{(1)}_0 \right)^2 + (h_1 -h_{\alpha})(h_1 - h_{\alpha} - 1 )\right) \tilde{G}_{N} (\mathbf{z}|h, \tilde h)  = 0, \\
\\
\left( \mathcal{L}^{(N)}_1 \mathcal{L}^{(N)}_{-1} + \mathcal{L}^{(N)}_0 - \left( \mathcal{L}^{(N)}_0 \right)^2 + (h_N -h_{\alpha})(h_N - h_{\alpha} - 1 )\right)  \tilde{G}_{N} (\mathbf{z}|h, \tilde h)  = 0,
\ea
\ee
thus, they are differential Casimir equations associated with external operators of dimensions $h_1 - h_{\alpha}$ and $h_N - h_{\alpha}$. Note that for the examples of $2,3$-pt necklace blocks analyzed above, there are no other Casimir equations that would be related to intermediate block dimensions. For $N>3$ the remaining $(N-3)$ Casimir equations ($j=3,..., N-1$) in \eqref{CAS} reduce to
\be
\left( \mathbb{L}_{1} \mathbb{L}_{-1} + \mathbb{L}_0 - \mathbb{L}^2_0 +  \tilde h_{j-2}(\tilde h_{j-2} - 1 )\right) \tilde{G}_{N} (\mathbf{z}|h, \tilde h)  =0,  \qquad \mathbb{L}_{P} = \sum^{N}_{k=j} \mathcal{L}^{(k)}_P,
\ee
where $\mathcal{L}^{(k)}_P$ are defined by \eqref{DO}. These equations represent the system of Casimir equations \cite{Rosenhaus:2018zqn, SimmonsDuffin:2012uy, Dolan:2011dv,Alkalaev:2022kal} associated with the set of intermediate dimensions $(\tilde h_{1},..., \tilde h_{N-3})$ of the comb channel block $\tilde{G}_{N} (\mathbf{z}|h, \tilde h)$.

We also check the Casimir equations for the $N$-pt blocks \eqref{generalNpt} with $s_{1,2} = 0, 1$. For these cases, the Casimir equations for $(s_1 +1) (s_2 +1)$ comb blocks $\tilde{G}^{(k,l)}_N (\mathbf{z}| h, \tilde h)$ in \eqref{generalNpt} are used in the proof. The check for larger $s_{1,2}$ is straightforward but it is not clear how to manifest that the system \eqref{CAS} is satisfied.

\vspace{-5mm}

\section{Conclusion}
\label{sec:cn}
 In this paper, we studied particular examples of the necklace blocks on the torus corresponding to the condition \eqref{mi}. We computed the $2$- and $3$-pt block functions for the first few $s_{1,2}$ which were found to be a product of the particular comb channel block and the factor which carries $q$ dependence. These results were generalized to the $N$-pt case where such a factorization takes place for the simplest case $s_{1,2}=0$. For all these necklace block functions, we explicitly checked that they satisfy the Casimir equations \eqref{CAS}. Namely, it was shown how the given system of equations for the $N$-pt necklace block is reduced to the Casimir equations for the particular $N$-pt comb channel block.

Initially, our study was motivated by the search for closed expression (in terms of known special functions) for necklace blocks using the condition \eqref{mi}. In  order to simply analysis one can relax the condition even more and consider degenerate primary operators with $h = - l/2, ~ l=1,2,...$ only. Then, the necklace block \eqref{maintool} contains a finite sum over $m$ and is a polynomial in the variable q. It also implies that the comb block reduces to a polynomial in $\chi_i$. We anticipate that for such cases it will be possible to find closed expressions for necklace blocks.

The obtained expressions for conformal blocks could help to explore the AdS/CFT correspondence between Wilson lines in the Chern-Simons 3d gravity theory and global torus blocks. Recently, it has been studied for the blocks with bosonic ($h=-j, \; j= 1, 2, ...$) degenerate operators \cite{Alkalaev:2020yvq}. Notice that for such operators the condition \eqref{mi} is obviously satisfied.  It would also be interesting to generalize these results to $\mathcal{N}=1$ superblocks by calculating them both explicitly and by analyzing as solutions of the Casimir equations. Another interesting direction is the analysis of these blocks and the Casimir equations \eqref{CAS} in the context of the Knizhnik-Zamolodchikov-Bernard equation \cite{Knizhnik:1984nr, BERNARD198877, BERNARD1988145}.

\vspace{4mm}

\textbf{Acknowledgements.} I am grateful to E. Akhmedov, K. Alkalaev, H. M. Babujian,  A. P.  Isaev, S. Mandrygin and R. H. Poghossian for useful discussions.  I also would like to thank the organizers of the XVIII International Conference on Symmetry Methods in Physics for the hospitality and productive atmosphere. The work was supported by the Russian Science Foundation grant No 18-72-10123.

\appendix

\section{Necklace blocks and OPE blocks}
\label{sec:nvo}

Here we discuss difference between OPE and necklace blocks focusing on a $2$-pt correlator. The 2-pt correlator is the first example which can be expand into two non-trivial channels corresponding to OPE and necklace blocks. Both types of blocks are basis functions in the space of $2$-pt correlators, but they have different properties. 

First, these blocks are expansions around different singular points (in $x$ coordinate), which can be easily seen from the series structure. More precisely, a 2-pt necklace block has the form
\be
\label{expl_2}
\ba{c}
\dps \cF(q, \mathbf{z}| h, \tilde h) = z^{-h_1}_1 z^{-h_2}_2 \sum^{\infty}_{n,m=0} \frac{\tau_{n,m}(\tilde h_1, h_1, \tilde h_2) \tau_{m,n}(\tilde h_2, h_2, \tilde h_1)}{n!m! (2\tilde h_1)_n (2 \tilde h_2)_m} q^{\tilde h_1  +n} x^{\tilde h_2 - \tilde h_1 + m- n}, \quad x = z_1/z_2, \\
\dps \tau_{m,n}(a,b,c) = m!n!\sum_{p=0}^{min[m,n]}\frac{(2c+n-1)^{(p)}(c+b-a)_{n-p}(a+b-c+p-n)_{m-p}}{p!(m-p)!(n-p)!}\,.
\ea
\ee
which is an expansion around $x=0$.  In turn, an OPE block is given by \cite{Kraus:2017ezw}
\be
\ba{c}
\dps \cF^{OPE}(q, \mathbf{z}| h, \tilde h) = \frac{ q^{h_{\alpha}}}{(1-q)^{\tilde{h}_1}} ~  _2F_1 (1-\tilde{h}_1, 2h_{\alpha} -\tilde{h}_1, 2 h_{\alpha}, q)
\\
\dps  \times  z^{-h_1}_1 z^{-h_2}_2 (1-x)^{\tilde h_1 - h_1 - h_2} {}_2F_1 \left( \tilde h_1, \tilde h_1 +h_{2}-h_{1}, 2 \tilde h_1 \big| 1 - x \right),
\ea
\ee
and here the hypergeometric function ${}_2F_1$ is an expansion defined for $|1-x|<1$. 

It is useful to compare these blocks on the torus with the $4$-pt blocks on the sphere. More precisely, $2$-pt OPE and necklace blocks can be constructed from $t$- and $s$- channel $4$-pt blocks on the sphere, respectively. These $4$-pt blocks are expressed in terms of the hypergeometric function ${}_2F_1$ and are connected by simple analytic continuations following from its properties. Despite this, for $2$-pt blocks such an analytic continuation is unknown and presumably has a much more complicated form, in particular, due to identification on the torus $q \sim q z$. It is also important to emphasize that on the torus OPE and necklace blocks have topologically non-equivalent diagrams.

In light of this, the expressions for necklace blocks carry information about singularities that cannot be extracted from OPE blocks. Moreover, as we will see, the expressions obtained in Section  \bref{sec:ebf} allow us to analyze the lines of singularities that occur in necklace blocks where the variables $z$ and the modular parameter $q$ are mixed. 

\section{More examples of the torus blocks in the necklace channel}
\label{mex}

\subsection{$2$-pt blocks for various $s$}
\label{sec:ma2}
Here we list expressions for several $2$-pt blocks corresponding to the case  $s \equiv h - \tilde h_0 - h_{\alpha} \neq 0$. As mentioned earlier, these blocks can be represented as
\be
\label{obsA}
\dps \cF^{(s)}_{2}(q, z_1, z_2|h, h_{\alpha}) = \dps \cF^{(0)}_{2}(q, z_1, z_2|h,\tilde h_0, h_{\alpha}) (1- q x)^{-2s} \tilde{P}^{(s)}(q, x),  \qquad x = z_1/z_2,
\ee
where
\be
 \tilde{P}^{(1)}(q, x) = \frac{h_{\alpha} q^2  x (2 \tilde h_0 + x)+q (\tilde h_0 x (x -2 -4 h_{\alpha})+\tilde h_{0}-2 h_{\alpha} x)+2 \tilde h_{0} h_{\alpha} x+h_{\alpha}}{2 \tilde h_0 h_{\alpha}}.
\ee
\be
\ba{c}
\tilde{P}^{(2)}(q, x) = \dps \frac{1}{2 \tilde h_{0}^2+\tilde h_{0}}+\frac{2 x}{\tilde h_{0}}+x^2 + \frac{q (2 \tilde h_{0} (x (x-4 h_{\alpha}-2)+1)+x (x-6 h_{\alpha}-2)+1)}{\tilde h_{0} (2 \tilde h_{0}+1) h_{\alpha}}\\
\\
+ \dps \frac{2 q x \left( x^2 -2 (h_{\alpha}+1) x+1\right)}{h_{\alpha}} + \frac{q (2 \tilde h_{0} (x (x-4 h_{\alpha}-2)+1)+x (x-6 h_{\alpha}-2)+1)}{\tilde h_{0} (2 \tilde h_{0}+1) h_{\alpha}}
\\
\\
\dps + \frac{q^2 x (4 \tilde h_{0}+x+2)}{\tilde h_{0} (2 \tilde h_{0}+1)} + \frac{q^2 \left(-8 (h_{\alpha}+1) x^3+2 (h_{\alpha} (6 h_{\alpha}+11)+7) x^2-8 (h_{\alpha}+1) x+x^4+1\right)}{h_{\alpha} (2 h_{\alpha}+1)}
\\
\\
\dps + q^4 x^2+\frac{2 q^3 x \left(-2 ( h_{\alpha}+1) x+x^2+1\right)}{ h_{\alpha}}.
\ea
\ee
In a similar manner one can find such polynomials for $s = 3, 4, ...$  but the general structure of such polynomials remains unclear for the general case. Starting from $s=3$, explicit functions become quite massive and we do not present them here. Nevertheless, we have verified \eqref{obsA} and \eqref{POLPR} for $s$ up to 4.  It is also straightforward to check that the blocks with given values of $s$ satisfy the Casimir system \eqref{CAS} for $N=2$.

\subsection{$3$-pt blocks for $s_{1,2} = 0, 1$}
\label{sec:ma3}
For $s_{1,2} = 0, 1$ we find that the $3$-pt necklace blocks are given by
\be
\cF^{(s_1,s_2)}_{3}(q, z|h,\tilde h, h_{\alpha}) = \dps \tilde{G}_3 (\mathbf{z}|\tilde h, h) (1- q y)^{-s_1 - s_2} P_3^{(s_1, s_2)}(q, x, y), \quad y = \frac{z_1}{z_3}, \quad x = \frac{z_1}{z_2}.
\ee
where $\tilde{G}_3 (\mathbf{z}|\tilde h, h)$ is the $3$-pt conformal block with external dimensions $(h_1-h_{\alpha}, h_2, h_3 - h_{\alpha})$. The first polynomials are
\be
\ba{c}
P_3^{(1, 0)}(q, x, y) = \left(h_2-h_3\right) (q y-1) + \\
\\
\dps \frac{-h_1 ((q-2) x y+q (y-2) y+x+y)+y (2 h_{\alpha} (q-1) (x-1)+(q-2) x-2 q+1)+q y^2+x}{x-y}.
\ea
\ee

\be
P_3^{(0, 1)}(q, x, y) = \frac{\left(h_{\alpha} -h_1+h_2\right) (q y-1)}{2 \left(1 -h_3+h_{\alpha}\right)}+\frac{q x (y-2)+(q-2) y+x+1}{2-2 x}.
\ee

\providecommand{\href}[2]{#2}\begingroup\raggedright\endgroup
\end{document}